\documentclass[10pt]{article}
\usepackage[T2A]{fontenc}
\usepackage[cp866]{inputenc}
\usepackage[english,russian]{babel}
\usepackage[tbtags]{amsmath}
\usepackage{amsfonts,amssymb,mathrsfs,amscd,amsthm}
\usepackage[lmargin=.75in,rmargin=.75in,tmargin=1.in,bmargin=1in]{geometry}
\theoremstyle{plain}

\theoremstyle{plain}

\theoremstyle{definition}

\def\({\left(}
\def\){\right)}

\def\bad{\spaceskip=0.33emplus0.6emminus0.15em\immediate\write5{\string\bad}}

\usepackage[dvips]{graphicx}

\begin{document}
\large

\vspace{0.25cm}
\begin{center}
\sf \Large
EXACT SOLUTIONS OF THE PROBLEM OF FREE-BOUNDARY\\ UNSTEADY FLOWS\\
\vspace{0.25cm} \rm
E. A. Karabut\\
 {\large Lavrentyev Institute of Hydrodynamics,
Lavrentyev av.,15, 630090
Novosibirsk\\  e-mail: eakarabut@gmail.com}\\

\end{center}

{\normalsize \vspace{0.5cm}
\noindent
Some approach to the solution of boundary value problems for finding functions, which are analytical in a wedge, is proposed. If the ratio of the angle at the wedge vertex to a number $\pi$ is rational, then the boundary value problem is reduced to the finite system of ordinary differential equations. Such approach, applied  the problem of inertial motion of a liquid wedge, made it possible to sum series with small denominators arising in the problem and find four exact examples of self-similar flows with a free boundary.}

\section{Statement of the problem}

Very few exact solutions of the problem of planar  unsteady flows of ideal incompressible fluid with free boundaries are known. One of the  examples is flows with a linear velocity field [1, 2].

\begin{center}
\resizebox{0.35\textwidth}{!}{\includegraphics{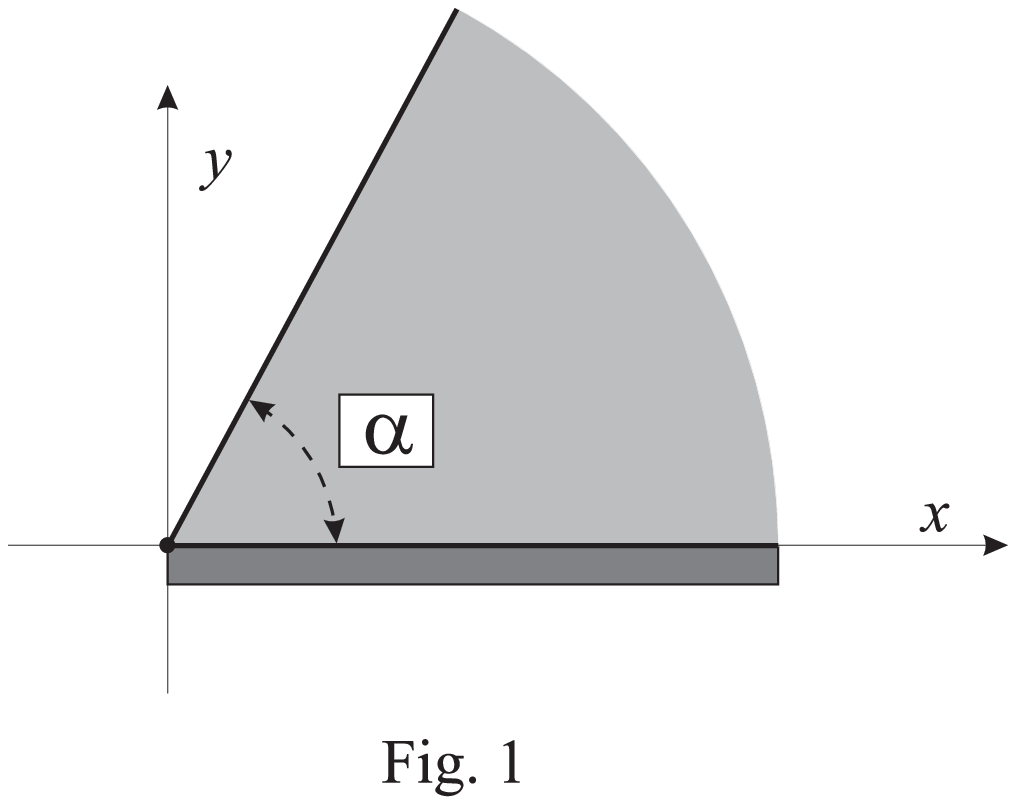}}
\end{center}

In the present work, new solutions  have been obtained. The following problem is considered. Initially, at  time $t=0$, the fluid  occupies the wedge with an apical angle $\alpha$ (Fig. 1). The origin of the Cartesian coordinate system $x$,~$y$ is placed at the vertex of the wedge. One boundary of the fluid is  a rigid fixed  wall $y = 0$, the other one is a free surface. Capillarity and  gravitation are absent.
 
The hard wall is added for the purposes of the symmetry.  If we remove it, then we have a liquid wedge with an  angle $2\alpha$. 

It is necessary to obtain the free surface and the complex velocity
$$U(z,t)=u(x,y,t)-{\rm i}v(x,y,t)$$ at later times  $t>0$. Here $z=x+{\rm i}y$ and  $u(x,y,t),\  v(x,y,t)$ are Cartesian velocity components. 

The fluid movement is caused by the initial velocity field, which is supposed to be quadratic:
$$U(z,0)=Az^2.$$

The problem contains only one dimensional real-valued parameter $A$.
Therefore, the problem is self-similar.   The solution of the problem in self-simulated variables  depends on only  the sign of this parameter.

\section{Power series in time}

We will seek a solution in the form of a power series in time. It follows from the dimension considerations that for the complex velocity, such series has the form:
\begin{equation}
U(z,t)=Az^2(1+a_1Azt+a_2(Azt)^2+a_3(Azt)^3+\ldots).
\label{s1}
\end{equation}
  We obtain from \eqref{s1}, that $U(0,t)=0$. This means that a particle of the liquid initially situated at the vertex of the wedge always remains  there. No  dimensionless function of time exists. Therefore, the apical angle is always constant.

 The impermeability condition with $y=0$ implies, that  the coefficients $a_j$ are supposed to be real-valued. All these coefficients can be obtained   from the pressure constancy   condition on the free surface as follows.
\begin{equation}\label{s2}
\begin{split}
a_1&=-\frac{2}{\cos 4\alpha},
\qquad
a_2=\frac{5}{2}\frac{3\cos\alpha-\cos 7\alpha}{\cos 4\alpha \cos 5\alpha},
\\
a_3&=\frac{-\frac{7}{2}\cos\alpha-5\cos 3\alpha-\frac{23}{2}\cos 5\alpha-
7\cos 7\alpha+\frac{7}{2}\cos 9\alpha+4\cos 11\alpha+\frac{23}{4}
\cos 13\alpha-\frac{1}{4}\cos 19\alpha}{\cos^2 4\alpha \,\cos 5\alpha\,\cos 6\alpha},\ldots\\
\end{split}
\end{equation}

Small denominators arise   in the solution. For example, at $\alpha=\pi/8$ acceleration and  pressure are initially infinite. It is not clear whether it has some physical sense or it was caused by  inadequacy of the series.

\section{Conformal mapping}
Formulae \eqref{s2} become more complicated very rapidly, however  series \eqref{s1} can be summed exactly for some values of $\alpha$.
In this paper, summation is performed  for three values of $\alpha$:  $\ \pi/4$, $\ \pi/2$, $\ 3\pi/4$. It turned out to be possible due to the usage of conformal mapping.

Let us consider in the auxiliary  plane $\zeta$ a wedge with an angle $\alpha$ and a vertex at  point $\zeta=0$. Let  $Z(\zeta,t)$ be a conformal mapping of this wedge onto the flow region and   $U(\zeta,t)$ be
a complex velocity. These  functions have the following representation:
$$Z=\frac{g(\zeta)}{At},\quad
U=\frac{f(\zeta)}{At^2}.
$$

Boundary conditions to find  unknown functions $f(\zeta), g(\zeta)$ are of the form [3]:
\begin{equation}\label{s3}
\begin{split}
{\rm Im}&\, g_r(\overline{g}+f)=0,\qquad
{\rm Re}\,(-2g_rf+f_rg+f_r\overline{f})=0,\qquad(\zeta=r{\rm e}^{{\rm i}\alpha})
\\
{\rm Im}&\, f=0,\qquad\qquad\quad\quad{\rm Im}\,g=0.\qquad\qquad\qquad\qquad\quad  ({\rm Im}\,\zeta=0)
\end{split}
\end{equation}

\section{Generating functions}
Let us consider the functions
\begin{equation}\label{s4}
P_j(\zeta)=g\left(\zeta\, {\rm e}^{2{\rm i}\alpha (j-1)}\right),
\quad
Q_j(\zeta)=f\left(\zeta\, {\rm e}^{2{\rm i}\alpha (j-1)}\right),
\end{equation}
which are the same functions $g(\zeta),\ f(\zeta)$ but they are in the coordinate system rotated by the angle $2\alpha(j-1)$ with respect to the initial one.
From  boundary conditions \eqref{s3} we obtain

{\bf Lemma 1.}
{\it Functions $P_j(\zeta),\  Q_j(\zeta)$ satisfy the infinite system  of ordinary differential equations
$$
\frac{{\rm d}P_{j+1}}{{\rm d}\zeta}
(P_j+Q_{j+1})=
\frac{{\rm d}P_{j}}{{\rm d}\zeta}
(P_{j+1}+Q_{j}),
$$
$$
2\frac{{\rm d}P_{j+1}}{{\rm d}\zeta}Q_{j+1}+
2\frac{{\rm d}P_{j}}{{\rm d}\zeta}Q_j=
\frac{{\rm d}Q_{j+1}}{{\rm d}\zeta}(P_{j+1}+Q_j)+
\frac{{\rm d}Q_j}{{\rm d}\zeta}(P_j+Q_{j+1}).
$$
}

If $\alpha/\pi$  is a rational number, then after some finite number of turns of the coordinate system we get the initial location of the system. Therefore from \eqref{s4} we have
\vspace{0.5cm}

{\bf Lemma 2.} {\it If\  $\alpha=\pi m/n$, where $m,n$ are natural numbers, and  $f(\zeta)$, $g(\zeta)$ are analytical functions in some vicinity of $\zeta=0$, then
$
P_{n+1}=P_1,\  Q_{n+1}=Q_1.
$
}

\vspace{0.5cm}

Two formulated lemmas imply the following

\vspace{0.5cm}

{\bf Theorem 1.}
{\it
If $\alpha=\pi m/n$, where $m,n$ are natural numbers, then to find functions  $f(\zeta)$ and $g(\zeta)$, which  are assumed by analytical  in some vicinity of $\zeta=0$, it is sufficient to solve the system of $2n$ ordinary differential equations
\begin{equation}\label{s5}
\begin{split}
\frac{{\rm d}P_{j+1}}{{\rm d}\zeta}
(P_j+Q_{j+1})&=
\frac{{\rm d}P_{j}}{{\rm d}\zeta}
(P_{j+1}+Q_{j}),\qquad(j=\overline{1,n})
\\
2\frac{{\rm d}P_{j+1}}{{\rm d}\zeta}Q_{j+1}+
2\frac{{\rm d}P_{j}}{{\rm d}\zeta}Q_j&=
\frac{{\rm d}Q_{j+1}}{{\rm d}\zeta}(P_{j+1}+Q_j)+
\frac{{\rm d}Q_j}{{\rm d}\zeta}(P_j+Q_{j+1}),\quad(j=\overline{1,n})
\\
P_{n+1}&=P_1,\qquad Q_{n+1}=Q_1
\end{split}
\end{equation}
 and then set
$
g=P_1,\  f=Q_1.
$
}

\section{Special cases}
Solving  system \eqref{s5} for $\alpha=\pi/2$, we obtain:
\begin{equation}\label{s6}
g(\zeta)=\zeta+\zeta^2,\quad f(\zeta)=\zeta^2, \quad
U(z,t)=\frac{\left(\sqrt{1+4zAt}-1\right)^2}{4At^2}.
\end{equation}
The complex velocity contains a singular  point of branching
$
x^*=-{1}/{(4At)}.
$
Since singular  points cannot be inside the liquid,  solution \eqref{s6} is valid only for $A > 0$. It is unknown, how to construct a solution for $\alpha=\pi/2$ at $A < 0$.

On the free surface $\zeta={\rm i}r$, where $r$ is real-valued. By using imaginary and real parts of the expression
$$
Z=({{\rm i}r+({\rm i}r)^2})/{(At)},
$$
we obtain the free surface form to be a parabola
\begin{equation}\label{s7}
xAt=-(yAt)^2.
\end{equation}

The evolution of \eqref{s7} with  increase of time at $A=1$ is shown at Fig. 2.
The dot denotes the location of the singularity $x^*$, which is initially  outside the liquid and finally it reaches the free surface. The initial half-plane  is transformed into the plane with rectilinear cut.

\begin{center}
\resizebox{0.7\textwidth}{!}{\includegraphics{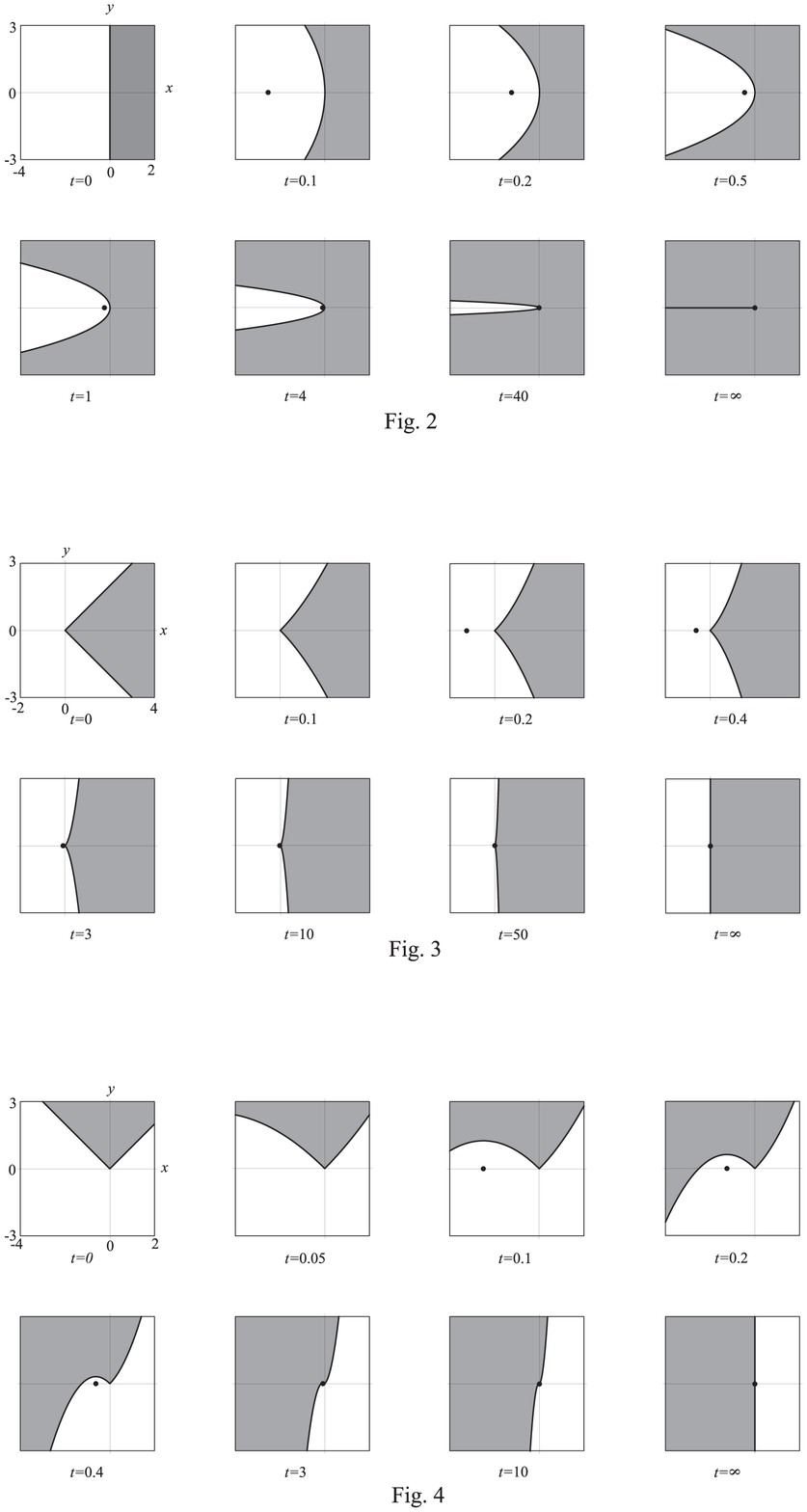}}
\end{center}

For both cases $\alpha=\pi/4$ and $\alpha=3\pi/4$, system \eqref{s5}, consisting of 8 equations, has the same form and yields the same solutions:
$$
g(\zeta)=\zeta-\zeta^2,\qquad
f(\zeta)=\zeta^2,\qquad
U(z,t)=\frac{\left(\sqrt{1-4zAt}-1\right)^2}{4At^2}.
$$
However, the free surface is described by different formulae. Substituting $\zeta=r{{\rm e}^{{\rm i}\pi/4}}$ and $\zeta=r{{\rm e}^{3{\rm i}\pi/4}}$ into the obtained solution, we get  the following forms of the free surface correspondingly:
\begin{equation}\label{s8}
yAt=xAt-2(xAt)^2,
\quad
yAt=-xAt+2(xAt)^2.
\end{equation}
The singular point of branching
$
x^{**}={1}/({4At})
$
should be outside liquid, therefore,
we suppose that $A<0$. It is unknown how to construct  a solution for $\alpha=\pi/4$ and $\alpha=3\pi/4$ at $A>0$.

The form of the free surface described by the first formula of \eqref{s8}, and the singular point $x^{**}$ for $A=-1$ are shown in Fig. 3.
The flow, where the liquid wedge with the right angle is transformed into a half-plane, was found.

The flow of the liquid located between two free surfaces \eqref{s8} for $A=-1$ is shown in Fig. 4.
The liquid wedge with the right apical angle, whose bisector is along the $y$ axis, evolves to a half-plane with increase of time.
Parameter $A$ can be both positive or negative for such flow because the singularity $x^{**}$ is anyway to be located outside the fluid.

A remarkable feature of the two flows,  shown in Figs. 3 and  4,  is the same form of the free surface.
Therefore, we can obtain the flow for $\alpha=3\pi/4$ if we join the upper half of the flow shown in Fig. 3 and the flow shown in Fig. 4.
As a result, we have the evolution of the fluid wedge with the angle $3\pi/4$ into the wedge with the angle $3\pi/2$.

\section{Conclusion}

The suggested technique makes it also possible to obtain new exact solutions for other rational $\alpha/\pi$.
Small denominators result from the resonance of the eigenfunctions of the linear boundary problem, which can be stated in a small vicinity of peak of the liquid wedge.

\section*{Acknowledgements}
The author thanks P.I. Plotnikov  for useful
discussions.

\end{document}